\documentclass[a4paper,12pt]{article}
\def \bsigma{\mbox{\boldmath $\sigma$}}

\begin{document}
\hspace{11cm} GEF-TH-1/2000
\vspace {1.5cm}

\begin{center}
\noindent {\bf On the miracle of the Coleman-Glashow and other 
baryon \linebreak[4] mass formulas} 

\vskip 30 pt

G.Dillon and G.Morpurgo
\vskip 7 pt
Universit\`a di Genova
and Istituto Nazionale di Fisica Nucleare, Genova, Italy.
\end{center}

\vskip 40 pt
\noindent {\bf Abstract.}
Due to a new measurement of the $\Xi^{0}$ mass, the Coleman-Glashow formula 
for the baryon octet e.m. masses (derived 
using unbroken flavor $SU_{3}$) is satisfied to an 
extraordinary level of precision. The same unexpected precision exists 
for the Gell Mann-Okubo formula and for its 
octet-decuplet extension (G. Morpurgo, Phys. Rev. Lett. 68 (1992) 139). 
We show that the old question ``why do they work so well?'' is now answered by 
the general parametrization method.
\\
({\em PACS:} 12.38.Aw; 13.40.Dk; 14.20.-c)

\vskip 40pt 

\baselineskip 24pt
\noindent {\bf 1. Introduction}
\vskip 5 pt

A recent measurement of the $\Xi^{0}$ mass \cite{ex} lowered considerably 
its error. The $\Xi^{0}$ mass is now $1314.82 \pm 0.06 \pm 0.2$ MeV. 
The importance of a new measurement was noted long ago \cite{em} in 
connection with the Coleman-Glashow (CG) e.m. mass formula. The 
previous value was $1314.9\pm 0.6$ MeV [3]. Indeed now the agreement 
of CG with the data is more miracolous than ever. Writing the CG 
formula as:
\begin{equation}
	p-n=\Sigma^{+}-\Sigma^- + \Xi^- -\Xi^{0}
	\label{CG}
\end{equation}
the present data give:
\begin{equation}
	l.h.s.=-1.29\ \mathrm{MeV} \qquad\qquad
r.h.s.=-1.58\pm 0.25 \ \mathrm{MeV}	\label{2}
\end{equation}
Because the mass differences $\Sigma^{-}-\Sigma^+$  
in (\ref{CG}) is $\approx 8$ MeV, the agreement is amazing ($\cong 
(4\pm 3)\%$) [before \cite{ex}, it was already excellent \cite{em} ($1.29$ 
to be compared with $1.67\pm 0.6$)]. To appreciate the point, note 
that the CG formula was derived \cite{CG} assuming unbroken flavor 
SU(3); but flavor is violated -in the baryon octet- by $\approx 
33\%$.

A similar situation applies to the Gell Mann-Okubo mass formula and 
its octet-decuplet extension by one of us \cite{mf}. It also holds -with 
larger errors- for some formulas of Gal and Scheck \cite{GS}. We 
already discussed \cite{em} all these relations using the QCD general 
parametrization method, but the result \cite{ex} suggests a 
revisitation. Indeed we are dealing perhaps with one of the most 
precise estimates in processes where the strong interactions play a 
role.

As stated above, the original derivation of CG neglected entirely 
the flavor breaking of the strong interactions. But it was shown in 
\cite{em} that the CG formula can be derived also keeping {\em all the 
flavor breaking terms}, with the only omission of terms with 3-quark 
indices. Here we complete the derivation \cite{em}; we include, in addition 
to the terms considered in \cite{em}, the effect of the $m_{d}-m_{u}$ mass 
difference and the so called Trace terms, absent in \cite{em}; they do not 
alter the conclusions of \nolinebreak[4] \cite{em}.
\vskip 40pt
\noindent {\bf 2. A brief summary of the general parametrization 
 method }
\vskip 5 pt
 
It is convenient to recall briefly the QCD parametrization method 
\cite{m, dm}. The method, based only on general features of QCD, 
applies to a variety of QCD matrix 
elements or expectation values. By integrating on all internal 
$q \bar{q}$ and gluon lines, the method parametrizes exactly such matrix 
elements. 
Thus hadron properties -like e.m. masses, including their 
flavor-breaking contributions- are written exactly as a sum 
of some spin-flavor structures each multiplied by a coefficient. Each 
structure (term) has, for baryons, a maximum of three indices. The 
coefficients of the various terms are seen to decrease with increasing 
complexity of the term. By the way this ``hierarchy" explains why the 
non relativistic quark model (NRQM), that keeps only the simplest 
terms, works quantitatively fairly well. Though the parametrization is 
performed in a given Lorentz frame and is, therefore non covariant, 
it is relativistic, being derived exactly from a relativistic field 
theory, QCD. For the basis of the method see \cite{m, dm}; 
applications are also given in [2,5,7-11]. Other references are listed in 
\cite{ pl, nc}; the latter gives a short review.

Here we will not recall the details of the method, but -for 
completeness- summarize it. 
The e.m. contribution to the mass of a baryon $B$ is:
\begin{equation}
	\langle \psi_{B}|\Omega | \psi_{B}\rangle = 
	\langle \phi_{B}|V^{\dag} \Omega V| \phi_{B}\rangle
	\label{3}
\end{equation}
In (3) $\Omega$ is -to second order in the charge- the exact QCD 
operator, expressed in terms of quark fields, representing the e.m. 
contribution to the mass, including all flavor breaking contributions 
of the strong interactions; $|\psi_{B}\rangle$ is the exact eigenstate  
of $B$ at rest 
of the QCD Hamiltonian; $| \phi_{B}\rangle$ is an auxiliary 
three body state of $B$, factorizable as
\begin{equation}
	|\phi_{B}\rangle=|X_{L=0}\cdot W_{B}\rangle
	\label{4}
\end{equation}
into a space part $X_{L=0}$ with orbital angular momentum zero and a 
spin unitary-spin part $W_{B}$. The unitary transformation $V$ 
-applied to the auxiliary state $|\phi_{B}\rangle$- transforms the 
latter into $| \psi_{B}\rangle$. After integration on the space 
variables, (4) can be written
\begin{equation}
	\langle \psi_{B}|\Omega | \psi_{B}\rangle = 
	\langle \phi_{B}|V^{\dag} \Omega V| \phi_{B}\rangle=
	\langle W_{B}| \sum_{\nu} t_{\nu}\Gamma_{\nu} (s,f) | W_{B}\rangle
	\label{5}
\end{equation}
where $\Gamma_{\nu}$ are operators depending only on the spin and 
flavor variables of the three quarks in $\phi_{B}$ and the $t_{\nu}$'s  
are a set of parameters. Of course \cite{m} the operator $V$ dresses the 
auxiliary state $|\phi_{B}\rangle$ with $q\bar{q}$ pairs and gluons 
and also introduces configuration mixing. Thus it generates the exact 
QCD eigenstate $|\psi_{B}\rangle$. It is the factorizability of 
$|\phi_{B}\rangle$ that allows the second step in Eq.(5), eliminating 
the space coordinates. The list of $\Gamma_{\nu}(s,f)$'s in 
(5) for the e.m. masses was given in \cite{em}. 
As stated, we will revisit this parametrization.
\vskip 40pt
\noindent {\bf 3. The corrections to the octet masses in the 
Coleman-Glashow formula}
\vskip 5 pt

The e.m. corrections to the octet baryon masses including flavor 
breaking analyzed in \cite{em} left out the $m_{d}-m_{u}$ contribution and
the Trace terms that, in this paper, we will include later 
(Sect. 4). For the moment we reanalyze the results of \cite{em}. There we 
called $\delta_{i}B$ the e.m. mass correction of baryon $B$ ($i=0,1,2$ 
refers to the order of the $s$-flavor breaking). Thus by $\delta_{0}B$ 
we meant (and mean) the e.m. correction neglecting all flavor 
breaking effects, $\delta_{1}B$ is the e.m. correction including only 
first order flavor breaking effects and so on for $\delta_{2}B$.

The quantities $\Gamma_{\nu}(s,f)$ that enter in the construction of 
$\delta_{i}B$'s are given in Eqs. (17-19) of \cite{em}. There is nothing to 
change in these equations except for a (trivial) point of notation. In 
\cite{em} the strange quark was called $\lambda$ (the non strange 
ones $\mathcal{N}$ 
and $\mathcal{P}$). Here we use the standard current notation $s$, $d$, $u$. We 
do this because (see \cite{dm}) we can select as we like the $q^{2}$ of the 
renormalization point of the quark mass and we now turn to the 
standard $q$ around $1$ GeV.  Thus the projectors $P^{\lambda}$, 
$P^{\mathcal{N}}$, $P^{\mathcal{P}}$ in \cite{em} are now written $P^{s}$, 
$P^{d}$, $P^{u}$.

The $\Gamma_{\nu}$'s in Eqs.(6,7) below are the same as those 
in Eqs.(17, 18) of \cite{em} (the sum symbol in 
each $\Gamma_{\nu}$ is defined in \cite{em}); we transcribe them here:

1) $\Gamma$'s of {\em zero order} in flavor breaking:
\begin{equation}
	\begin{array}{lclcl}
			\Gamma_{1}=\sum [Q_{i}^{2}] & ; & \Gamma_{2}=\sum 
			[Q_{i}^{2}(\bsigma_{i} \cdot \bsigma_{k})] & ; & \Gamma_{3}=\sum 
			[Q_{i}^{2}(\bsigma_{j}\cdot \bsigma_{k})]  \\
			\Gamma_{4}=\sum [Q_{i}Q_{k}] & ; & \Gamma_{5}=\sum 
			[Q_{i}Q_{k}(\bsigma_{i}\cdot\bsigma_{k})] & ; & \Gamma_{6}=\sum 
			[Q_{i}Q_{k}(\bsigma_{i}+\bsigma_{k})\cdot\sigma_{j}]
		\end{array}
	\label{6}
\end{equation}

2) $\Gamma$'s of {\em first order} in $P^s$ (acting in 
$\Lambda$,$\Sigma$,$\Sigma^\ast$,$\Xi$,$\Xi^\ast$,$\Omega$):
\begin{equation}
	\begin{array}{lclcl}
			\Gamma_{7}=\sum [Q_{i}^{2}P_{i}^s] & ; & \Gamma_{8}=\sum 
			[Q_{i}^{2}P_{i}^s(\bsigma_{i}\cdot\bsigma_{k})] & ; & \Gamma_{9}=\sum 
			[Q_{i}^{2}P_{i}^s(\bsigma_{j}\cdot\bsigma_{k})]  \\
			\Gamma_{10}=\sum 
			[Q_{i}^{2}P_{k}^s] & ; & \Gamma_{11}=\sum 
			[Q_{i}^{2}P_{k}^s(\bsigma_{i}\cdot\bsigma_{k})] & ; & \Gamma_{12}=\sum 
			[Q_{i}^{2}P_{k}^s(\bsigma_{i}+\bsigma_{k})\cdot\bsigma_{j})]  \\
			\Gamma_{13}=\sum 
			[Q_{i}Q_{k}P_{i}^s] & ; & \Gamma_{14}=\sum 
			[Q_{i}Q_{k}P_{i}^s(\bsigma_{i}\cdot\bsigma_{k})] & ; & \Gamma_{15}=\sum 
			[Q_{i}Q_{k}P_{i}^s(\bsigma_{i}+\bsigma_{k})\cdot\bsigma_{j}]  \\
			\Gamma_{16}=\sum 
			[Q_{i}Q_{k}P_{j}^s] & ; & \Gamma_{17}=\sum 
			[Q_{i}Q_{k}P_{j}^s(\bsigma_{i}\cdot\bsigma_{k})] & ; & \Gamma_{18}=\sum 
			[Q_{i}Q_{k}P_{j}^s(\bsigma_{i}+\bsigma_{k})\cdot\bsigma_{j}]
		\end{array}
	\label{7}
\end{equation}
where $Q_{i}=\frac{2}{3}P_{i}^u - \frac{1}{3}P_{i}^d - 
\frac{1}{3}P_{i}^s$ and the $P_{i}$'s are the projectors on the $u$, 
$d$, $s$ quarks.

As to the $\Gamma_{\nu}$'s of second and third order in $P^s$, they 
are listed in Eqs. (19,20) of \cite{em}. We will not transcribe them here.

The expression $\delta_{0}B$ of the electromagnetic mass of $B$ at 
zero order in flavor breaking is:
\begin{equation}
	\delta_{0}B=a\Gamma_{1}+b\Gamma_{2}+c\Gamma_{3}+
	d\Gamma_{4}+e\Gamma_{5}+f\Gamma_{6}
	\label{8}
\end{equation}
(where, to agree with \cite{em}, we used $a \ldots f$ instead of $t_{1} 
\ldots t_{6}$). As shown in \cite{em} one can check that:
\begin{displaymath}
	\delta_{0}P-\delta_{0}N=\delta_{0}\Sigma^{+}-\delta_{0}\Sigma^{-}+
	\delta_{0}\Xi^{-}-\delta_{0}\Xi^{0}
\end{displaymath}
which is the CG relation at zero order in flavor breaking.

As to $\delta_{1}B$ and $\delta_{2}B$, coming from the first and 
second order in flavor breaking, we refer to \cite{em}. There it is 
shown (table III) that, except for terms with three indices, also 
$\delta_{1}B$ and $\delta_{2}B$ leave the CG relation unaltered. 
Thus, to {\em all orders in flavor breaking, with the only omission of 
three quark terms in $\delta_{1}B$ and $\delta_{2}B$}, the CG relation 
holds:
\begin{equation}
	\delta P -\delta N = \delta \Sigma^{+}-\delta \Sigma^{-}+\delta 
	\Xi^{-}-\delta \Xi^{0}
	\label{9}
\end{equation}
where now $\delta \equiv \delta_{0}+\delta_{1}+\delta_{2}$.

To evaluate the order of magnitude of the three quark terms of 
the type $\delta_{1}B$ and $\delta_{2}B$ (the only ones that violate 
the CG formula) we now use the hierarchy discussed at length in 
previous papers \cite{hi, dm}. Of course, the dominant order of 
magnitude, is here that of the two index terms of the form 
$Q_{i}Q_{k}$. Let us 
consider the three quark terms in $\delta_{1}B$. One can show that 
$\Gamma_{9}$, $\Gamma_{12}$ and $\Gamma_{15}$ do not contribute to the 
left and right hand sides of the CG formula. As to $\Gamma_{16}$, 
$\Gamma_{17}$ and $\Gamma_{18}$, they do not contribute to 
$\delta_{1}n$ and $\delta_{1}p$, while for 
$\delta_{1}\Sigma^{+}-\delta_{1}\Sigma^{-}+\delta_{1}\Xi^{-}-\delta_{1}\Xi^{0}$ 
one gets in all cases $(``something")/3$. As to the magnitude of 
$ ``something"$, note that according to the hierarchy, each of 
$\Gamma_{16}$, $\Gamma_{17}$ and $\Gamma_{18}$ carries a reduction 
factor $\approx 0.3$ due to the presence of a $P^{s}$ and $\approx 
0.3$ due to one more gluon exchange. Thus we have a reduction of 
order $(1/3)^{3}\cong 4\cdot 10^{-2}$ for each of the above terms with 
respect to the dominant no flavor-breaking contributions. (Note: 
$\Gamma_{18}$ here is $1/2$ that in \cite{em} due to an incorrect 
normalization in \cite{em} that produced, however, no effect because 
3-index terms were not evaluated in \cite{em}.)

Because experimentally $\Sigma^{-}-\Sigma^{+}\cong 8$ MeV and $ 
\Xi^{-}-\Xi^{0}\cong 6.4$ MeV, the above value of $4\cdot 10^{-2}$ 
implies an expected difference between left and right hand sides of 
the CG formula of $\approx 0.2 \div 0.3$ MeV that does not disagree 
with the data.

A similar argument holds for the three-quark terms of second order in 
flavor breaking listed as (19) in \cite{em}. These second order flavor breaking 
three-quark terms, are expected from the hierarchy, to be $\approx 
(1/9)$ of the first order flavor breaking terms mentioned above; that is
\begin{displaymath}
	\vert \delta_{2}\Xi^{-}-\delta_{2}\Xi^{0}\vert \approx 
	\frac{1}{9}\vert 
	\delta_{1}\Sigma^{+}-\delta_{1}\Sigma^{-}+
	\delta_{1}\Xi^{-}-\delta_{1}\Xi^{0}\vert
\end{displaymath}
and therefore the three quark, second order flavor breaking effect 
should contribute to the difference between left and right hand side 
of the CG relation by $\approx 0.02$ MeV.
\vskip 40pt
\noindent {\bf 4. The effects of $m_{d}-m_{u}$ and of the Trace terms 
on the CG relation}
\vskip 5 pt

a) {\em The effect of  $m_{d}-m_{u}$.} As shown in \cite{dm} the 
quantity that intervenes in evaluating the effect of a mass 
difference $\Delta m$ between quarks is $\Delta m/(\beta\Lambda)$ 
where $\Lambda \equiv \Lambda_{QCD} \cong 200$ MeV and $\beta$ is 
approximately 3. Because $\Delta m$ for $d$ and $u$ is a few MeV, 
only the first order term in the expansion in $\Delta m/(\beta 
\Lambda)$ may be relevant. But it is easy to check that 
the CG formula is not modified. (Obviously product terms of
 $\Delta m$ and $Q_{i}Q_{k}$ type perturbations are totally negligible.)

b) {\em The effect of the Trace terms.} In addition to the terms in 
\cite{em} other terms are present in the general parametrization 
(see \cite{dm}, in particular footnote 14). They leave the CG formula 
unaltered, as we will show; they must however be recorded. These 
``Trace'' terms correspond to QCD Feynman closed loops, as 
exemplified, e.g. in \cite{zs} (fig.1) or \cite{nc} (fig.3). 
Introducing the matrix
\begin{equation}
	Q=\frac{2}{3}P^u - \frac{1}{3} P^d - \frac{1}{3} P^s
	\label{10}
\end{equation}
(not to be confused with the baryon charge $Q_{B}$ that was called $Q$ 
in \cite{em}), the Trace terms can be constructed as follows: Consider 
the quantities
\begin{displaymath}
	T_{1}=Tr(QP^s) \ , \qquad T_{2}=Tr(Q^2) \ , \qquad T_{3}=Tr(Q^{2}P^{s})
\end{displaymath}
and multiply them by $\sum (\bsigma_{i}\cdot \bsigma_{k})$ or by the 
sums listed in Eqs.(6) and (7) keeping only those expressions that, 
after the multiplication, contain two $Q$ symbols (either $Q^{2}$ or 
$Q\cdot Q_{i}$ or $Q_{i}Q_{k}$) and a number of $P^{s}$ from 0 to 2; 
for instance  (just to exemplify)
\begin{equation}
	\begin{array}{lll}
			 Tr(Q^{2})\sum (\bsigma_{i}\cdot \bsigma_{k})\ ,
			 & Tr(Q^{2}P^{s}) \sum (\bsigma_{i}\cdot \bsigma_{k})\ ,
			 &	Tr(QP^{s})\sum Q_{i} \ ,
			 \\
			 Tr(QP^{s})\sum Q_{i}(\bsigma_{i}\cdot \bsigma_{k}) \ , 
			& Tr(QP^{s})\sum Q_{j}(\bsigma_{i}\cdot \bsigma_{k}) &
		\end{array}
	\label{11}
\end{equation}
It is easy to check that none of the above terms changes the previous 
conclusions concerning the exactness of the CG equation. The only new 
quantity that enters produced by terms of type (11) is the 
expectation value of $\sum_{i}Q_{i}\sigma_{zi}$ on the baryons.
 For instance $Tr(QP^{s}) 
\sum Q_{i}(\bsigma_{i}\cdot \bsigma_{k})=-Q_{B}/2 
+(3/2)\sum_{i}Q_{i}\sigma_{zi}$; the values of $\langle 
\sum_{i}Q_{i}\sigma_{zi}\rangle$ are listed in Table I of \cite{m}, 
indicated there as $\langle \Sigma_{z}^{q}\rangle$. One checks 
immediately that the CG equation remains true.

We conclude this discussion of the CG miracle as follows: The three 
quark terms are expected to give a very small, but non zero 
contribution to CG. It is remarkable that the hierarchy typical of 
the general parametrization, appears to explain this smallness thus 
providing one of those few cases where one can estimate a tiny effect 
of the strong interactions and find it compatible with the data.
\vskip 40pt
\noindent {\bf 5. The Gell-Mann Okubo mass formula and its extension 
to the octet-decuplet}
\vskip 5 pt

We will discuss now, first qualitatively, then quantitatively, the 
reason why, besides the CG formula, also the Gell-Mann 
Okubo (GMO) formula has its share of mysterious perfection. The GMO 
formula for the octet baryons:
\begin{equation}
	\frac{1}{2}(n+\Xi^{0})=\frac{1}{4}(3\Lambda +\Sigma)
	\label{12}
\end{equation}
is derived neglecting terms of second order in flavor breaking. The 
expansion parameter for flavor breaking is -as generally known and 
determined from the general parametrization \cite{m, dm}- $0.3$ to 
$0.33$. Because the magnitude of first order flavor breaking is
$\approx 150$ to $190$ MeV, an estimate of the contribution
of second order flavor breaking terms -neglected in (12)- is
$\approx 50$ to $60$ MeV. Instead the r.h.s and l.h.s  of (12) 
differ by $\approx 6$ MeV. Is this success just luck? Not entirely.
In fact, by the general parametrization, we can now parametrize 
together the decuplet and octet masses and determine in this way all the 
8 parameters in the octet+decuplet mass formula. The coefficient of 
the second order flavor breaking, $(a+b)$ in \cite{mf,dm} is in fact, 
as expected, $\approx 3$ times smaller then the above estimate $50$ or $60$ 
MeV, due to the fact that necessarily second order flavor breaking 
terms have two indices. 
Thus one expects a difference of $\cong 17$ or $20$ MeV 
between the left and right hand sides of (12). Because \cite{mf} the 
parameter that enters in the GMO formula is $(a+b)/2$ we are led, 
by this order of magnitude argument, to a difference of $8-10$ MeV, not 
far from the experimental value of $\approx 6$ MeV. The essential role 
is once more played by the hierarchy, that again leads to negligible 
three index second order flavor breaking terms, precisely as for the CG 
formula. 

Let us now be more quantitative. The fact that the coefficient $c$, $d$ 
(Eq. (4,5) of \cite{mf}) 
multiplying the second order flavor three index terms are 
negligibly small was first discovered in relation to the GMO formula 
\cite{mf} (see also \cite{dm}) and used above in discussing the CG formula.
Barring these $c$, $d$ coefficients one finds a mass formula that relates 
the octet and decuplet masses, correct to second order in flavor, 
except for a three index term. This formula is just 
\cite{mf} the GMO formula plus a ``decuplet" correction $T$:
\begin{equation}
	\frac{1}{2}(n+\Xi^{0})+T=\frac{1}{4}(3\Lambda 
	+2\hat{\Sigma}^{+}-\Sigma^{0})
	\label{13}
\end{equation}
where $\hat{\Sigma}^{+}\equiv 2\Sigma^{+}-\Sigma^{0}+2(p-n)$ and $T$ is:
\begin{equation}
	T=\Xi^{\ast -}-\frac{1}{2}(\Omega+\Sigma^{\ast -})
	\label{14}
\end{equation}
The charge specification are inserted here because, at this accuracy, 
Eq.(13) must take into account the e.m. contributions: The 
combination (13) is unaffected by the e.m. corrections if the latter 
are calculated at zero order flavor breaking. (Note: Eq.(13) is more 
simply $\frac{1}{2}(p+\Xi^{0})+T=\frac{1}{4}(3\Lambda +2\Sigma^{+}-\Sigma^{0})$ but 
here we kept the form used in \cite{mf}.)

While of course the improved $\Xi^{0}$ \cite{ex} slightly decreases 
the experimental error in (13), the main part of the error comes from 
the decuplet masses in $T$. With the conventional values of the masses 
\cite{pdg} it is:
\begin{displaymath}
	T(conventional)=5.18\pm 0.66 \ \mathrm{MeV}
	\end{displaymath}
whereas, if the pole values \cite{pdg} of the resonances are taken 
\cite{dm}, it is:
\begin{displaymath}
	T(pole)=6.67\pm 1.25 \ \mathrm{MeV}
\end{displaymath}
The left and right hand side of (13) become:
\begin{displaymath}
	\begin{array}{cll}
			 (conventional)& Left=1132.36\pm 0.7\ \mathrm{MeV} & 
			 Right=1133.93\pm 0.04\ \mathrm{MeV} \\
			(pole) & Left=1133.86\pm 1.25\ \mathrm{MeV} & 
			Right=1133.93\pm 0.04\ \mathrm{MeV}
		\end{array}
\end{displaymath}
In both cases it is again true that three quark terms breaking flavor to 
second order are estimated to contribute to the 
difference between r.h.s. and l.h.s. less than $0.7$ MeV.

We finally note that, to first order in $|m_{u}-m_{d}|/(\beta 
\Lambda_{QCD})$, the $u$, $d$ mass difference does not affect the 
octet-decuplet mass formula (13); also the Trace terms do not modify 
the parametrization in this case.
\vskip 40pt
\noindent {\bf 6. Other e.m. mass formulas}
\vskip 5 pt

Using the NRQM Gal and Scheck \cite{GS} derived long ago a set of 
relations among the electromagnetic masses for mesons and baryons. For 
mesons their assumptions were very restrictive, but for baryons they 
amounted mostly to the neglect of three body terms. In \cite{em} we 
showed how these formulas could be reproduced by the general 
parametrization method under certain assumptions. Although it might be 
of interest to reanalyze the situation in more detail, we refrain from 
it because the experimental data did not change substantially.

However, in order to stimulate more precise measurements (if possible), 
we write down below a relation that is totally analogous, for the 
decuplet, to the Coleman Glashow formula for the octet and that can be 
treated in exactly the same way. It is 
independent of the Gal Scheck derivation, but easily verifiable using 
the Eqs.(27, 28) of \cite{em}:
\begin{equation}
	\delta \Delta^{+}-\delta\Delta^{0}=\delta\Sigma^{\ast +}
	-\delta\Sigma^{\ast -}+\delta\Xi^{\ast -}-\delta\Xi^{\ast 0}
	\label{15}
\end{equation}
Incidentally Eq.(15), together with the relation (16) below, derived, 
as well known, from isospin algebra
\begin{equation}
	\delta \Delta^{++}-\delta\Delta^{-}=3(\delta 
	\Delta^{+}-\delta\Delta^{0})
	\label{16}
\end{equation}
might be of interest in connection with the determination of the 
mass differences between the $\Delta$'s.
More generally, Eq.(15) can be derived, using (16), from the charge 
corrected second order Okubo equations (7, 8) of ref.\cite{d}.
\vskip 40 pt
\noindent {\bf 7. Conclusion}
\vskip 5 pt

The reason for the extraordinary perfection of the Coleman-Glashow 
relation that holds at present to $\cong (0.29\pm 0.25)/8\cong (4\pm 3)\cdot 
10^{-2}$ (in spite of having been derived, we recall, using unbroken 
flavor SU(3)) is now clear. It depends, we have shown, on the 
smallness of the three index terms in the general parametrization. We 
underline again (Sect.4) that this smallness represents one of the few 
cases where, thanks to the hierarchy in the parametrization, an 
estimate of an effect due to the strong interaction can be given and 
found to be tiny as expected.

As to the Gell-Mann Okubo formula, its octet decuplet extension \cite{mf} 
including second order flavor breaking except for three index terms, 
holds to better than $2\cdot 10^{-2}$. This value is the ratio between 
the experimental error  ($\approx 1$ MeV)  and what 
one  would expect estimating the orders of magnitude ($\approx 50$ MeV) 
just by flavor breaking to second order. This  confirms the smallness of the 
three index terms. 
\pagebreak


\begin{thebibliography}{99}
\small
\baselineskip 20pt

\bibitem{ex} NA 48 Collab., V. Fanti et al., Eur. Phys. J. C12 (2000) 
69.

\bibitem{em} G. Morpurgo, Phys. Rev. D 45  (1992) 1686.

\bibitem{pdg} Particle Data Group, C. Caso et al., Eur. Phys. J. C 3 
(1998) 1.
 
\bibitem{CG} S. Coleman and S.L. Glashow, Phys. Rev. Lett. 6 
(1961) 423.

\bibitem{mf} G. Morpurgo, Phys. Rev. Lett.  68 (1992) 139.

\bibitem{GS} A. Gal and F. Scheck, Nucl. Phys. B 2 (1967) 121.

\bibitem{m} G. Morpurgo, Phys. Rev. D 40 (1989)  2997.
 
\bibitem{dm}  G. Dillon and  G.Morpurgo, Phys. Rev. D 53 (1996) 3754.

\bibitem{hi} G. Morpurgo, Phys. Rev. D 46 (1992) 4068.

\bibitem{zs} G. Dillon and G. Morpurgo, Zs. Phys. C 64 (1994) 467.

\bibitem{pl} G. Dillon and G. Morpurgo, Phys. Lett. B 459 (1999) 321.

\bibitem{nc} G. Morpurgo, La Rivista del Nuovo Cimento 22 (1999) 1.

\bibitem{ok} S. Okubo, Phys. Lett. 4 (1963) 14.

\bibitem{d} G. Dillon, Europhys. Lett. 20 (1992) 389.
 

\end{thebibliography}
\end{document}